\documentclass[twocolumn,showpacs,amsmath,amssymb,aps,pra,floatfix,nofootinbib,superscriptaddress]{revtex4}
\usepackage{graphicx,graphics,times,color}
\usepackage{bm}
\usepackage{longtable}
\usepackage{color}
\usepackage{lipsum}

\def\be{\begin{equation}}
\def\ee{\end{equation}}
\def\ba{\begin{eqnarray}}
\def\ea{\end{eqnarray}}
\def\la{\langle}
\def\ra{\rangle}

\def\h{\hskip 1cm}

\begin{document}

\title{Arbitrary perfect state transfer in $d$-level spin chains}

\author{Abolfazl Bayat}
\affiliation{Department of Physics and Astronomy, University College London, Gower St., London WC1E 6BT, United Kingdom}

\date{\today}

\begin{abstract}
We exploit a ferromagnetic chain of interacting $d$-level ($d>2$) particles for arbitrary perfect transfer of quantum states with $(d-1)$ levels. The presence of one extra degree of freedom in the Hilbert space of particles, which is not used in encoding, allows to achieve perfect transfer even in a uniform chain through a repeated measurement procedure with consecutive single site measurements. Apart from the first iteration, for which the time of evolution grows linearly with the size of the chain, in all other iterations, the evolution times are short and does not scale with the length. The success probability of the mechanism grows with the number of repetitions and practically after a few iterations the transfer is accomplished with a high probability.
\end{abstract}


\maketitle

\section{Introduction}

The natural time evolution of strongly correlated many-body systems can be exploited for propagating information across distant sites in a chain \cite{bose-review,bayat-review-book}. Very recently, experimental realization of such phenomena have been achieved in NMR \cite{state-transfer-NMR}, coupled optical wave-guides \cite{Nikolopoulos-perfect-transfer,kwek-perfect-transfer} and cold atoms in optical lattices \cite{Bloch-spin-wave,Bloch-magnon}. The idea has been generalized  for higher spins in both ferromagnetic \cite{bayat-dlevel-2007} and anti-ferromagnetic \cite{Sanpera-spin1} regimes.

In the simplest case of a uniform chain the quality of transport goes down with increasing the size of the system due to natural dispersion of excitations.
To achieve  perfect state transfer across a chain one idea is to engineer the Hamiltonian to have a linear dispersion relation (see Ref.~\cite{Kay-review} for a detailed review on perfect state transfer). This can be achieved by either engineering the couplings \cite{christandl} or local magnetic fields \cite{perfect-transfer-magnetic}. The engineered chains may also be combined with extra control on the boundary to avoid state initialization for perfect transfer \cite{DeFranco-perfect}.  One may also engineer the two boundary couplings \cite{leonardo} of free fermionic systems in order to excite only those eigenvectors which lie in the linear zone of the spectrum and thus achieve an almost perfect transfer. A sinusoidal deriving of the couplings \cite{hanggi} has also been proposed for routing information in a network of spins between any pair of nodes. A set of pulses in a system with both ferromagnetic and anti-ferromagnetic couplings \cite{Kay,Karimipour-perfect} may also be used to properly transfer quantum states across a spin network. An alternative for achieving almost perfect state transfer is engineering the spectrum of the system to create resonance between sender and receiver sites by either using very weak couplings \cite{weak-coupling} or strong local magnetic fields \cite{Perturbation-magnetic}.

As mentioned above, to achieve perfect state transfer, most of the proposed mechanisms are based on engineering the Hamiltonian of the system, up to some degrees, in order to, at least approximately, achieve linear dispersion relation or bring the sender and receiver in to resonance. In Ref.~\cite{burgarth-dual} a dual rail system with uniform couplings has been used in an iterative procedure to achieve arbitrary perfect state transfer of a qubit. This idea then has been generalized to multi-rail systems \cite{burgarth-multirail} for transferring higher level states as well. Although, in both dual and multi-rail systems the chains are uniform and no engineering is needed, but the price which is paid is the number of chains which are needed as well as the more complex encoding and decoding of the quantum states.

In this paper, a mechanism is introduced for arbitrary perfect state transfer, which has the same spirit of iterative procedure of the dual and multi-rail systems \cite{burgarth-dual,burgarth-multirail} but with much less complexity. According to this proposal, a ferromagnetic chain of interacting $d$-level ($d>2$) particles are used for sending quantum states with $(d-1)$ levels. The natural time evolution of the system with consecutive single particle projective measurement in a ceratin basis allows for iterative perfect state transfer. The probability of success grows with the number of iterations and apart from the first iteration, all others can be done within a very short time scale, which does not scale with size, and thus the time needed to achieve perfect transfer remains reasonable, even for long chains.

The structure of the paper is as following: In section \ref{sec2} the model is introduced, in section \ref{sec3} the mechanism for state transfer is discussed and in section \ref{sec4} the further iterations for achieving perfect transfer are analyzed. Finally, in section \ref{sec5} the results are summarized and a possible realization in optical lattices is discussed..

\begin{figure} \centering
    \includegraphics[width=8cm,height=6cm,angle=0]{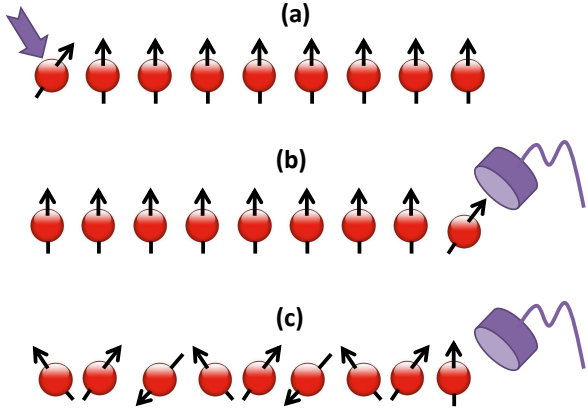}
    \caption{ (color online) (a) A ferromagnetic chain of interacting particles for quantum state transfer. The first spin is encoded in the state $|\phi_s\ra$ while the rest are prepared in the state $|0\ra$. (b) When the measurement is successful the quantum state is perfectly transferred to site $N$ and all the other spins reset to the state $|0\ra$. (c) When the measurement is unsuccessful the last site is projected in $|0\ra$ and the excitation is dispersed along the chain. }
     \label{fig1}
\end{figure}

\begin{figure*} \centering
    \includegraphics[width=15cm,height=5cm,angle=0]{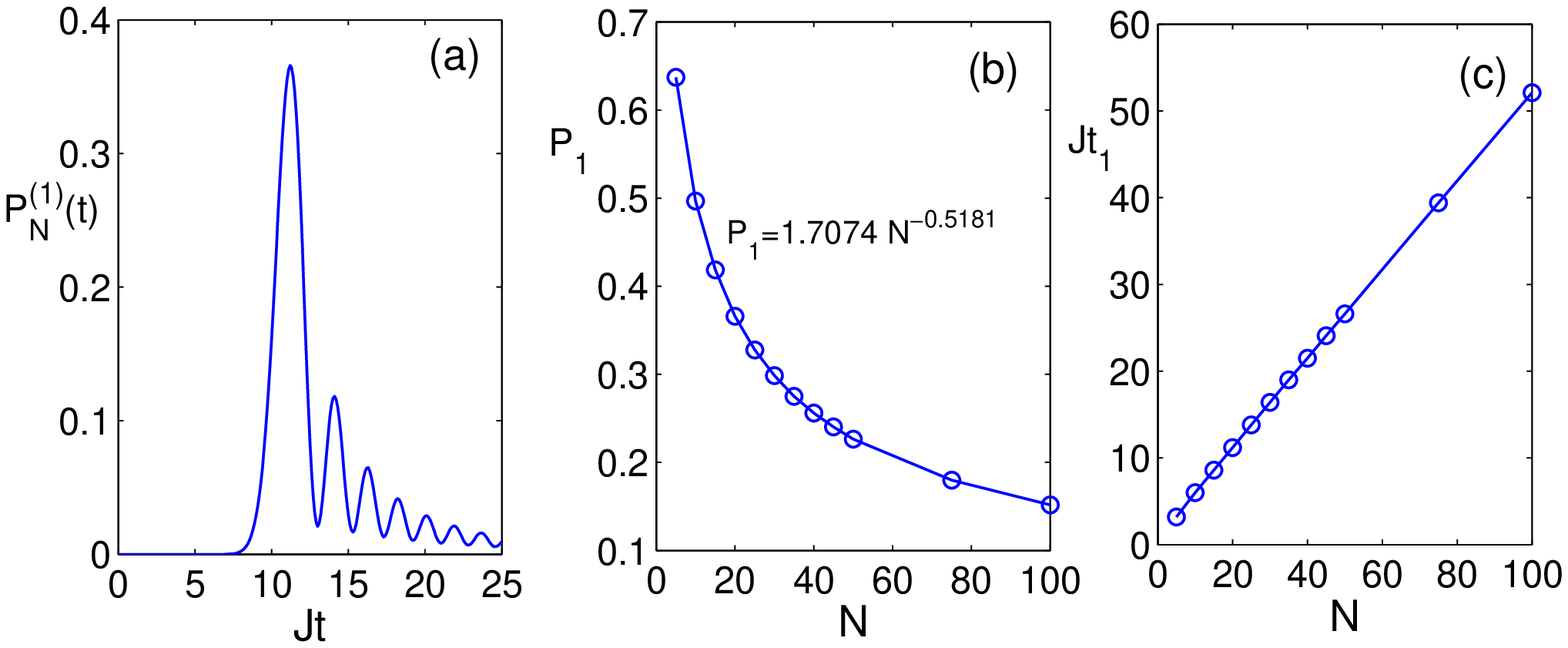}
    \caption{(Color online) (a)  The probability of success in the first iteration as a function of $Jt$ in chain of length $N=20$. (b) The probability of success $P_1$ in terms of length $N$. (c) The optimal time $Jt_1$ versus length $N$.  }
     \label{fig2}
\end{figure*}

\section{Introducing the model} \label{sec2}

We consider a chain of $N$ particles that each takes $\mu=0,1,...,d-1$ (for $d\geq 3$) different levels. They interact through the Hamiltonian
\begin{equation}\label{H}
    H=-J\sum_{k=1}^{N-1}  \mathbf{P}_{k,k+1}+B\sum_{k=1}^{N}S^z_k
\end{equation}
where $J$ is the exchange coupling, $B$ is the magnetic field, $\mathbf{P}_{k,k+1}$ is the swap operator which exchanges the quantum states of sites $k$ and $k+1$, and $S^z_k$ is the generalized Pauli operator in the $z$ direction acting on site $k$ which is defined as $S^z|\mu\ra=\mu|\mu\ra$. In the case of $J>0$ and vanishing magnetic field (i.e. $B=0$) the system is ferromagnetic and its ground state is $d$-fold degenerate of the form $|\mu,\mu,...,\mu\ra$. The two terms in the Hamiltonian commute with each other and thus, the effect of the magnetic filed $B>0$ is just to lift the degeneracy and choose the quantum state $|\mathbf{0}\ra=|0,0,...,0\ra$ as the unique ground state of the system. This makes it possible to initialize the system through a simple cooling procedure. After initialization, the magnetic field can be switched off as it has no effect on transport of an excitation, which will be seen in the following.

The  above Hamiltonian is one possible generalization of spin-$1/2$ Heisenberg interaction for higher spins \cite{bayat-dlevel-2007}.
In fact, the swap operator $\mathbf{P}_{k,k+1}$, for $d$-level systems, can always be written as
\begin{equation}\label{Swap_spin_operators}
    \mathbf{P}_{k,k+1}=\sum_{p=0}^{d-1} b_p (\mathbf{S}_k.\mathbf{S}_{k+1})^p
\end{equation}
where, $b_p$'s are some real numbers. To determine these coefficients we apply both sides of the above identity on general states of the form $|\mu \nu\ra$ for which we get $d(d+1)/2$ different equations. However, only $d$ equations are independent which are associated to the cases that the difference $|\mu-\nu|=k$ takes the values of $k=0,1,...,d-1$. By solving these $d$ independent equations one can uniquely determine all the $b_p$ coefficients. For instance, in the special case of spin-1 ($d=3$) one can easily show that
\begin{equation}\label{H2}
   \mathbf{P}_{k,k+1}= -I+\mathbf{S}_k.\mathbf{S}_{k+1}+(\mathbf{S}_k.\mathbf{S}_{k+1})^2
\end{equation}
where $I$ is the identity operator.

The Hamiltonian $H$ has several symmetries with corresponding conserved charges which includes
\begin{equation}\label{Qm}
    [H,Q^{(m)}]=0, \h Q^{(m)}=\sum_{k=1}^{N} (S^z_k)^m,
\end{equation}
for $m=1,2,...,d-1$. These set of conservation laws imply for example that a state like $|2,0,0,...,0\ra$ cannot evolve to a state like $|1,1,0,0,...,0\ra$. That is why we use the particular form of the Hamiltonian in Eq.~(\ref{H}) which is essential to achieve arbitrary perfect state transfer in our system.

The states with only one site excited are called one particle states and are represented as $|\mathbf{\mu}_k\ra=|0,0,...0,\mu,0,...,0\ra$ in which site $k$ is in state $|\mu\ra$ and the rest are in state $|0\ra$. The one particle sector with $Q^{(1)}$ charge equal to $\mu$ is denoted by $V_1^{(\mu)}$ and the whole one particle sector is
\begin{equation}\label{V_1_sector}
    V_1=V_1^{(1)}\oplus V_1^{(2)} \oplus ... \oplus V_1^{(d-1)}.
\end{equation}

The Hamiltonian $H$ in Eq.~(\ref{H}) can be analytically diagonalized in the $V_1$ subspace using Bethe ansatz. The eigenvectors and the corresponding eigenvalues are
\begin{eqnarray}\label{eigen_E}
    |E_\mu^m\ra&=&\sqrt{\frac{4}{2N+1}} \sum_{k=1}^{N} \sin(\frac{ 2m+1 }{2N+1}k \pi) |\mu_k\ra \cr
    E_\mu^m &=& \mu B-2J \cos(\frac{ 2m+1}{2N+1}\pi)
\end{eqnarray}
where $m=0,1,...,N-1$ and an irrelevant constant number has been dropped from the eigenvalues.

\section{Quantum state transfer} \label{sec3}

System is initially prepared in its ground state $|\mathbf{0}\ra$. The quantum state which has to be transferred is then encoded on the sender site $s$ (which is assumed to be 1 throughout this paper) in a general state as the following
\begin{equation}\label{psi_s}
    |\phi_s\ra=\sum_{\mu=1}^{d-1} a_\mu |\mu\ra,
\end{equation}
where $a_\mu$ are complex coefficients with the normalization constraint of $\sum_{\mu=1}^{d-1} |a_\mu|^2=1$. It has to be emphasized that this general state does not include $\mu=0$ (which all other spins are prepared to) and thus the dimension of its Hilbert space is $d-1$. This extra degree of freedom in the chain will then be used to achieve the {\em perfect} transfer of the quantum state in Eq.~(\ref{psi_s}). A schematic picture of the system when the initialization is accomplished is shown in Fig.~\ref{fig1}(a) in which the sender spin $1$ is prepared in quantum state $|\phi_s\ra$ while the rest are all in state $|0\ra$. Thus, the initial state of the system becomes
\begin{equation}\label{psi_0_1}
    |\Psi^{(1)}(0)\ra=\sum_{\mu=1}^{d-1} a_\mu |\mathbf{\mu}_1\ra.
\end{equation}
Since the excitation is located in site $1$ this quantum state is not an eigenstate of the system and hence the system evolves under the action of the Hamiltonian
\begin{equation}\label{psi_t_1}
    |\Psi^{(1)}(t)\ra=e^{-iHt}|\Psi^{(1)}(0)\ra.
\end{equation}

Using the symmetries of Eq.~(\ref{Qm}) One can easily show that
\begin{equation}\label{psi_tt_1}
    |\Psi^{(1)}(t)\ra=\sum_{\mu=1}^{d-1} \sum_{k=1}^N a_\mu f_{k1}^\mu(t) |\mathbf{\mu}_k\ra,
\end{equation}
where,
\begin{equation}\label{f_nm_mu}
    f_{nm}^\mu(t)=\la \mathbf{\mu}_n|e^{-iHt}|\mathbf{\mu}_m\ra
\end{equation}
form a unitary $N\times N$ matrix $f^\mu(t)$ with the elements given in Eq.~(\ref{f_nm_mu}).
Using the eigenvectors of the Eq.~(\ref{eigen_E}) together with the fact that in the subspace $V_1$ we have $\sum_{m=0}^{N-1}\sum_{\mu=1}^{d-1} |E_\mu^m\ra \la E_\mu^m|=I$ one can show that $f^\mu(t)=e^{-i\mu B t}F(t)$ where the elements of matrix $F(t)$ are given as
\begin{widetext}
\begin{equation}\label{f_nm_mu2}
    F_{mn}(t)=\frac{4}{2N+1}\sum_{p=0}^{N-1} e^{i2Jt\cos(\frac{ 2p+1}{2N+1}\pi)} \sin(\frac{ 2p+1}{2N+1}m \pi) \sin(\frac{  2p+1 }{2N+1}n\pi).
\end{equation}
\end{widetext}

To extract the quantum state at a receiver site $r$ one measures the following operator at site $r$
\begin{equation}\label{O_measurement}
    O_r=\sum_{\mu=1}^{d-1} |\mu\ra \la \mu|.
\end{equation}
If the outcome of measurement is $1$ (i.e. measurement is successful) then the quantum state is perfectly (up to a local unitary rotation) transferred to site $r$ as schematically shown in Fig.~\ref{fig1}(b). Otherwise, if the outcome of measurement is $0$ (i.e. measurement is unsuccessful) the spin at site $r$ is projected to state $|0\ra$ and the excitation is dispersed across the chain as schematically shown in Fig.~\ref{fig1}(c). The probability of a successful measurement on site $N$ (the receiver site $r$ is assumed to be $N$ throughout this paper) at time $t$ is
\begin{equation}\label{Psuc_1}
    P_N^{(1)}(t)=\la \Psi^{(1)}(t)| O_N|\Psi^{(1)}(t) \ra = |F_{N1}(t)|^2
\end{equation}

In Fig.~\ref{fig2}(a) we plot $P_N^{(1)}(t)$ as a function of time $Jt$ in a chain of length $N=20$. As it is clear from the figure the probability peaks for the first time at a particular time $t=t_1$ and then its oscillations damp over time. To maximize the probability of success we should thus perform the measurement at time $t=t_1$ which gives the probability of success as
\begin{equation}\label{P_1_t1}
    P_1=P_N^{(1)}(t_1)=|F_{N1}(t_1)|^2
\end{equation}

In Fig.~\ref{fig2}(b) the success probability $P_1$ is plotted in terms of length $N$ which is perfectly fit by $P_1=1.7074N^{-0.5181}$. The optimal time $Jt_1$ is also plotted as a function of $N$ in Fig.~\ref{fig2}(c) which shows a perfect linear dependence on the length. In the case of success, the quantum state at the receiver site $N$ becomes
\begin{equation}\label{psi_r1_local}
    | \phi_N^{(1)} \ra = \sum_{\mu=1}^{d-1} a_\mu e^{-i\mu B t_1}|\mu\ra
\end{equation}
which is different from $|\phi_s\ra$, given in Eq.~(\ref{psi_s}). To convert this state into $|\phi_s\ra$ one has to perform a local unitary operator of the form $e^{iBS_zt_1}$ into site $N$ which then accomplishes the perfect state transfer.

\section{Further Iterations} \label{sec4}

The quantum state transfer, discussed above, is not always successful as the probability of success $P_1$ is less than 1 and decreases by increasing the length $N$. In fact, if the measurement is unsuccessful the quantum state of the whole chain is projected into
\begin{equation}\label{psi_0_2}
    |\Psi^{(2)}(0)\ra=\frac{1}{\sqrt{1-P_1}}\sum_{\mu=1}^{d-1} \sum_{k=1}^{N-1} a_\mu f^\mu_{k1}(t_1)|\mathbf{\mu}_k\ra
\end{equation}
where the index $k$ runs from $1$ to $N-1$ because the previous measurement had been unsuccessful and thus the receiver site $N$ is inevitably projected to state $|0\ra$. In addition, the reason that the parameter of  $|\Psi^{(2)}(0)\ra$ is chosen to be $0$ is due to the fact that we start another iteration now and the system has not yet evolved in this part. A very interesting feature is revealed by exploring the spatial distribution of excitations across the chain in this quantum state by computing
\begin{equation}\label{P_n_distribution}
    P_m^{(2)}(0)=\la \Psi^{(2)}(0)| O_m|\Psi^{(2)}(0) \ra=\frac{|F_{m1}(t_1)|^2}{1-P_1}
\end{equation}
where index $m$ takes $m=1,2,...,N-1$.

\begin{figure} \centering
    \includegraphics[width=9cm,height=4.5cm,angle=0]{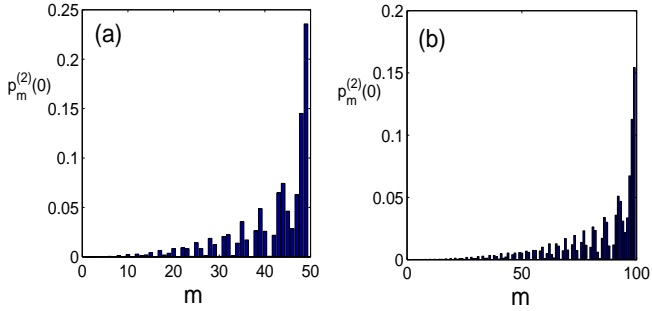}
    \caption{(Color online) The distribution of excitations across the chain when the first measurement is unsuccessful for a chain of length (a) $N=50$ and; (b) $N=100$.}
     \label{fig3}
\end{figure}

\begin{figure} \centering
    \includegraphics[width=9cm,height=6cm,angle=0]{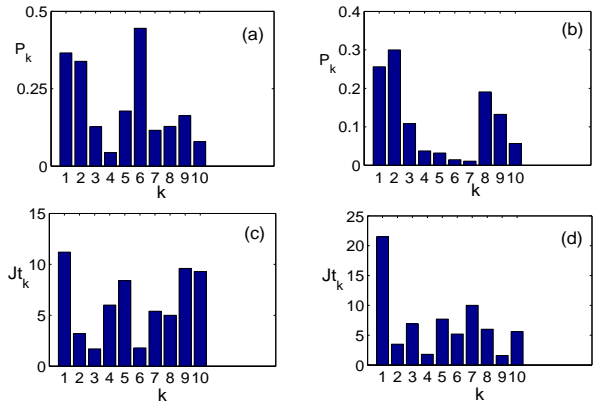}
    \caption{(Color online) The probability of success in each iteration $k$ in chains of length: (a) $N=20$; (b) $N=40$. The corresponding optimal times $t_k$ for each iteration $k$ in chains: (c) $N=20$; (d) $N=40$.}
     \label{fig4}
\end{figure}

In Figs.~\ref{fig3}(a)-(b) we plot the distribution function $P_m^{(2)}(0)$ in terms of site index $m$ for two different chains of length $N=50$ and $N=100$ respectively. As it is evident from the figures the distribution is more prominent near the end of the chain, in particular at $m=N-2$ and $m=N-1$. This is indeed due to the particular optimization of $t_1$ in the first iteration which is chosen to maximize the probability of receiving the excitation at site $N$ and thus in the case of unsuccessful measurement the excitations are still very close to the last site of the chain. System is free to evolve as
$|\Psi^{(2)}(t)\ra=e^{-iHt}|\Psi^{(2)}(0)\ra$ and just as before one has to perform the measurement of Eq.~(\ref{O_measurement}) on site $N$ and see if the outcome is $1$. The probability of success in this iteration is
\begin{equation}\label{Psuc_2}
    P_N^{(2)}(t)=\la \Psi^{(2)}(t)| O_N|\Psi^{(2)}(t) \ra.
\end{equation}
In contrast to the first iteration, to maximize this probability we no longer need to wait for very long times as according to the distribution $P_m^{(2)}(0)$, shown in Figs.~\ref{fig3}(a)-(b), the excitations are very close to the receiver site $N$. Indeed, the time window, over which the optimization is done, for all iterations after the first one can be fixed and independent of $N$. In this paper, all time optimizations for iterations after the first one are taken in a time interval of $0\leq Jt\leq10$.
The results are, in fact, hardly improved by choosing a wider time window. At a particular time $t=t_2$ at which the probability $P_N^{(2)}(t)$ peaks the measurement is performed on site $N$. Hence, the success probability is $P_2=P_N^{(2)}(t_2)$ which can be written as
\begin{equation}\label{Psuc_2}
    P_2=\frac{1}{1-P_1} |\sum_{m=1}^{N-1} F_{Nm}(t_2) F_{m1}(t_1)|^2.
\end{equation}
In the case of successful measurement the quantum state of site $N$ becomes
\begin{equation}\label{psi_r2_local}
    | \phi_N^{(2)} \ra = \sum_{\mu=1}^{d-1} a_\mu e^{-i\mu B (t_1+t_2)}|\mu\ra
\end{equation}
which then is converted into the target state $|\phi_s\ra$ by applying the local unitary operator $e^{-iBS_z(t_1+t_2)}$.

\begin{figure} \centering
    \includegraphics[width=8cm,height=5cm,angle=0]{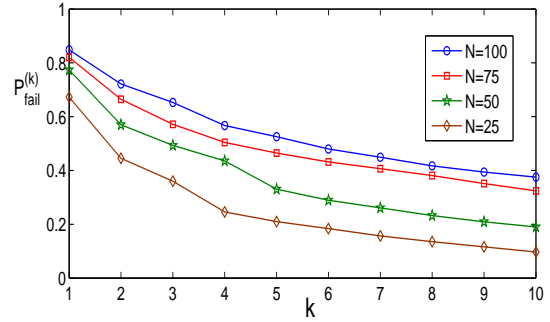}
    \caption{(Color online) The probability of failure after $k$ consecutive iterations versus the number of iterations $k$ in chains of different lengths.}
     \label{fig5}
\end{figure}

In the case of unsuccessful measurement we can repeat the process over and over till the quantum state reaches the receiver site. One can easily show that after $k$ unsuccessful iterations the quantum state of the whole system at the beginning of the $(K+1)$'th iteration is
\begin{widetext}
\begin{equation}\label{psi_0_k}
    |\Psi^{(k+1)}(0)\ra=\frac{1}{\sqrt{(1-P_{k})(1-P_{k-1})...(1-P_1)}}
    \sum_{\mu=1}^{d-1} \sum_{m_k=1}^{N-1}...\sum_{m_1=1}^{N-1}
    a_\mu f^\mu_{m_{k} m_{k-1}}(t_{k}) ... f^\mu_{m_2 m_1}(t_2) f^\mu_{m_11}(t_1) |\mathbf{\mu}_k\ra.
\end{equation}
\end{widetext}

Then system is released to evolve just as before, i.e.  $|\Psi^{(k+1)}(t)\ra=e^{-iHt}|\Psi^{(k+1)}(0)\ra$. The probability of a successful measurement after time $t$ is $P_N^{(k+1)}(t)=\la \Psi^{(k+1)}(t)| O_N|\Psi^{(k+1)}(t) \ra$
from which the time $t_{k+1}$ is determined as its maximum in the time interval of $0\leq Jt \leq 10$ and thus we have $P_{k+1}=P_N^{(k+1)}(t_{k+1})$. One can show that

\begin{widetext}
\begin{equation}\label{psi_0_k}
    P_{k+1}=\frac{1}{(1-P_{k})(1-P_{k-1})...(1-P_1)}
   | \sum_{m_k=1}^{N-1}...\sum_{m_1=1}^{N-1}
    F_{Nm_k}(t_{k+1}) F_{m_k m_{k-1}}(t_{k}) ... F_{m_2 m_1}(t_2) F_{m_1 1}(t_1) |^2
\end{equation}
\end{widetext}

It is worth mentioning that in the case of successful measurement the received quantum state is
\begin{equation}\label{psi_r_k_local}
    |\phi_N^{(k+1)}\ra =\sum_{\mu=1}^{d-1} a_\mu e^{-i\mu B \sum_{m=1}^{k+1} t_m} |\mu\ra
\end{equation}
which can be converted to $|\phi_s\ra$ by the local unitary operation of $e^{iBS_z\sum_{m=1}^{k+1} t_m}$. In the case of unsuccessful measurement the process is repeated again.

In Figs.~\ref{fig4}(a)-(b) we depict $P_k$ for different iterations $k$ in chains of length $N=20$ and $N=40$ respectively. As it is clear from the figures in some iterations the probability of success is relatively small which is due to the limitation that we used for the time window in which the optimization is performed. In Figs.~\ref{fig4}(c)-(d) the corresponding optimal times $Jt_k$ is shown for each iteration on the same chains. As shown in the figure, apart from $t_1$ all other times are less than $10/J$ which was imposed to the system by us to avoid long waiting times.

It is worth mentioning that the above choice of $t_k$'s are just to maximize the probability of success in the shortest possible time. In fact, the algorithm works for any choice of $t_k$ including the regular waiting times of $t_k=(2k-1)t_1$ used in Ref.~\cite{burgarth-dual} which represents the oscillation  of the excitation along the chain due to reflection from the boundaries.

The best way to see the performance of our mechanism is to compute the probability of failure after $k$ consecutive iterations. This means that the procedure has to fail in all $k$ iterations whose probability is then
\begin{equation}\label{psi_r_k_local}
    P_{fail}^{(k)}=\prod_{m=1}^k (1-P_m).
\end{equation}

In Fig.~\ref{fig5} we plot $P_{fail}^{(k)}$ as a function of iteration $k$ for different chains. As it is evident from the figure the probability of failure goes down by increasing the iterations. For instance, in a long chain of length $N=100$ after $10$ iterations the probability of failure is $\sim 0.35$ while for smaller chains the situation is of course much better as, for example, in a chain of length $N=25$ after $10$ iterations the probability of failure is less than $0.1$.

\section{Realization in Optical Lattices}

Cold atoms are the most promising candidate for realizing $d$-level chains. A Bose-Hubbard chain \cite{Jacsh-bose-hubbard} containing  $^{87}$Rb or $^{23}$Na atoms in the half-filling regime can be tuned to its Mott insulator phase, where exactly one atom recites in each site. For such system the interaction between the atoms can be explained by an effective spin-1 Hamiltonian as \cite{Yip-spin1-singlet}
\begin{equation}\label{Heff_spin1}
    H=\sum_{n=1}^{N-1} J \mathbf{S}_n.\mathbf{S}_{n+1}+ K (\mathbf{S}_n.\mathbf{S}_{n+1})^2
\end{equation}
where $J=-\frac{2t^2}{U_2}$ and $K=-\frac{2t^2}{3U_2}-\frac{4t^2}{3U_0}$ for which $t$ is the tunneling and $U_S$ ($S=0,2$) is the on-site interaction energy for the two spin-1 particles with the total spin $S$. By tuning $U_2=U_0$ one gets $J=K$ and thus the swap operator of Eq.~(\ref{H2}) is realized for spin-1 atoms. The quantum phases accessible to the ground state \cite{GS-cold-spin1} and the quench dynamics \cite{Quench-cold-spin1} of $S=1$ spinor atoms have already been analyzed.

Higher level atoms have also been investigated to realize spin-2 ($d=5$) \cite{Cold-spin2} and spin-3 ($d=7$) \cite{Cold-spin3} spinor gases which can be used to realize our proposed mechanism. Although, tuning the interaction to be of the form of swap operators of Eq.~(\ref{H}) might be tricky and needs more detailed analysis of the hyperfine levels and the interaction of atoms in such chains.

\section{Conclusion} \label{sec5}

We proposed a mechanism for transferring a quantum state of $(d-1)$-levels across a $d$-level spin chain ($d>2$) through an iterative procedure which its probability of success grows continually with the number of iterations. The fidelity of the transferred quantum state is perfect, up to a local unitary rotation, while the Hamiltonian is uniform and no complicated engineering of the couplings is needed. In the process, apart from the first iteration, for which the optimal time $t_1$ grows linearly with the length $N$, in all subsequent iterations the evolution time is short and does not grow with the size of the system. This is very useful as in the presence of decoherence the fast operation time does not allow the environment to spoil the quality of state transfer that much. Furthermore, when the transfer is accomplished the system is automatically rests to its initial ferromagnetic state and becomes ready for reusing. The proposed mechanism is most suited for realization in the fast growing field of trapped atoms in optical lattices.

In compare to dual \cite{burgarth-dual} and multi-rail \cite{burgarth-multirail} systems our proposed mechanism is simpler for fabrication as it only needs a single spin chain, no matter how large $d$ is. In contrast, for sending quantum states of larger $d$ one has to increase the number of chains in the multi-rail systems which then makes both the encoding and decoding processes very complicated. In addition, it is worth mentioning that in dual rail systems the excitations are {\em delocalized} between different parallel spin chains which makes it vulnerable against dephasing.   \\

{\em Acknowledgements:-} Discussions with Sougato Bose, Leonardo Banchi, Matteo Scala and Enrico Compagno are warmly acknowledged. This paper was supported  by the EPSRC grant $EP/K004077/1$.

\end{document}